\documentclass[twocolumn,superscriptaddress]{revtex4}
\usepackage{graphicx}

\begin{document}

\title{Unlocking the full potential of wave-matter nonlinear coupling in the epsilon-near-zero regime}

\author{Alessandro Ciattoni}
\affiliation{Consiglio Nazionale delle Ricerche, CNR-SPIN, Via Vetoio 10, 67100 L'Aquila, Italy}

\author{Carlo Rizza}
\affiliation{Dipartimento di Scienza e Alta Tecnologia, Universit\`a dell'Insubria, Via Valleggio 11, 22100 Como, Italy}
\affiliation{Consiglio Nazionale delle Ricerche, CNR-SPIN, Via Vetoio 10, 67100 L'Aquila, Italy}

\author{Andrea Marini}
\affiliation{ICFO-Institut de Ciencies Fotoniques, Mediterranean Technology Park, 08860 Castelldefels (Barcelona), Spain}

\author{Andrea Di Falco}
\affiliation{SUPA, School of Physics and Astronomy, University of St. Andrews, North Haugh, St. Andrews KY16 9SS, United Kingdom}

\author{Daniele Faccio}
\affiliation{School of Engineering and Physical Sciences, SUPA, Heriot-Watt University, Edinburgh EH14 4AS, United Kingdom}

\author{Michael Scalora}
\affiliation{Charles M. Bowden Research Center RDMR-WDS-WO, RDECOM, Redstone Arsenal, Alabama 35898-5000, USA}

\begin{abstract}
In recent years, unconventional metamaterial properties have triggered a revolution of electromagnetic research which has unveiled novel scenarios of wave-matter interaction. A very small dielectric permittivity is a leading example of such unusual features, since it produces an exotic static-like regime where the electromagnetic field is spatially slowly-varying over a physically large region. The so-called epsilon-near-zero metamaterials thus offer an ideal platform where to manipulate the inner details of the "stretched" field. Here we theoretically prove that a standard nonlinearity is able to operate such a manipulation to the point that even a thin slab produces a dramatic nonlinear pulse transformation, if the dielectric permittivity is very small within the field bandwidth. The predicted non-resonant releasing of full nonlinear coupling produced by the epsilon-near-zero condition does not resort to any field enhancement mechanisms and opens novel routes to exploiting matter nonlinearity for steering the radiation by means of ultra-compact structures.
\end{abstract}

\maketitle

During the last decade the metamaterial route for achieving unusual electromagnetic properties has attracted a great deal of interest in both theoretical and applied research, since it has brought to light novel electromagnetic regimes \cite{Valent,Pendr1,Chennn,Poddub,Mahoud}, and has suggested a number of remarkable devices for extreme manipulation of the radiation \cite{Pendr2,Liuuu1,Liuuu2}. Structures exhibiting very small dielectric permittivity, or epsilon-near-zero (ENZ) metamaterials \cite{Rizza1,Maasss,Gaooo1,Sunnn1,Ciatt1}, belong to the family of media able to affect electromagnetic radiation in a very unconventional way because the medium's effective wavelength is much larger than the vacuum wavelength, and because they host a regime where both field amplitude and phase are slowly-varying over relatively large portions of the bulk, which is quite opposite to geometrical optics. Such key feature has been exploited to conceive setups where ultra-narrow ENZ channels are able to "squeeze" electromagnetic waves at will \cite{Silve1,Silve2,Edwar1,Liuuu3}, and to develop new paradigms of devices for tailoring the antenna radiation pattern \cite{Aluuu1,Luoooo,Soricc}. In addition, ENZ metamaterials have also been shown to support a rich phenomenology of surface waves \cite{Ciatt2,Vassan,Travis,Ciatt3,Liiiii,Campi1,Newman}, to achieve perfect absorption \cite{Zhongg}, to enhance spatial dispersion effects \cite{Pollar}, and to support novel cloaking mechanisms \cite{Bilott,Liznev}.

Other interesting mechanisms and effects arise when the ENZ regime is combined with matter nonlinearity. A kind of crucial interplay between small linear permittivity and optical nonlinearity has been identified in Ref.\cite{Husako}, where the authors discuss an all-optical transition from dielectric to metal behavior in a nonlinear ENZ slab. A similar mechanism has been exploited to predict a class of solitons \cite{Ciatt4} where the intensity-driven dielectric-metal transition occurs along the transverse soliton profile, yielding exotic features like transverse power flow reversing \cite{Ciatt5}, unusual shapes like hollow-core \cite{Rizza2}, and two-peaked and flat-top profiles \cite{Rizza3}. Furthermore, the combination of the ENZ regime with nonlinearity benefits from the non-resonant enhancement of the normal electric field component across the vacuum-ENZ medium interface \cite{Campi2}, producing intriguing effects like transmissivity directional hysteresis \cite{Ciatt6,Ciatt7} and enhancement of second and third harmonic generation  \cite{Vince1,Ciatt8,Capret,Lukkkk}. A different field enhancement mechanism has been identified within narrow ENZ plasmonic channels, which has been exploited to boost optical nonlinearities \cite{Argyr1}, to investigate temporal soliton excitation \cite{Argyr2}, and for the enhancement of second-harmonic generation efficiency \cite{Argyr3}.

\begin{figure*}
\center
\includegraphics[width=1\textwidth]{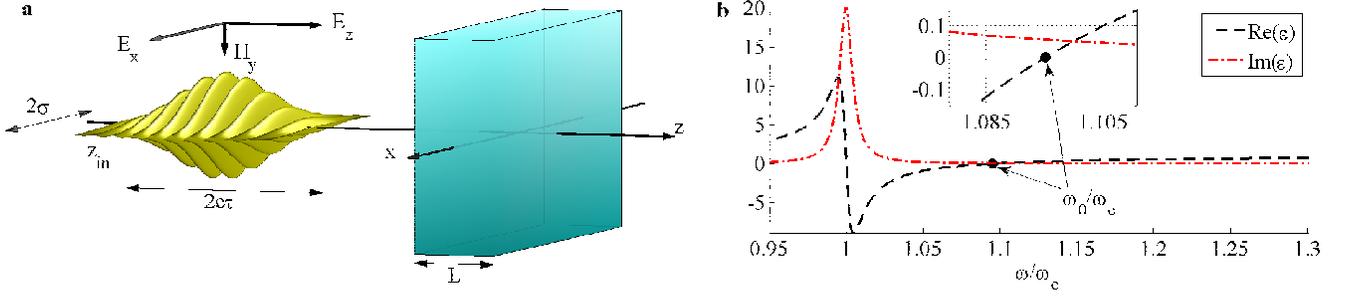}
\caption{\textbf{Pulse scattering by a slab supporting the ENZ regime}. (\textbf{a}) Geometry of the scattering interaction between a transverse magnetic pulse which is both spatially and temporally localized and a material slab having, in the linear regime, zero-crossing-points of the real part of its dielectric permittivity. (\textbf{b}) In the linear regime where the medium polarization $|\textbf{P}|$ is much smaller than the saturation polarization $P_s$ (see Eq.(\ref{polariz})), the slab has a dielectric permittivity $\epsilon (\omega)$ (see Eq.(\ref{eps})) with a standard Lorentz profile located at the resonant frequency $\omega_e$ and with a zero-crossing-point of its real part at $\omega_0$ (see Eq.(\ref{omega0})). Dispersion parameters have been chosen in such a way that the imaginary part of the permittivity is low around zero-crossing-point so that $|\epsilon(\omega)|$ is much smaller than one in a spectral bandwidth around $\omega_0$ and the slab can support the ENZ regime.}
\end{figure*}

In this work we theoretically prove that the interplay between the ENZ condition and the optical nonlinearity triggers a novel nonlinear matter-wave coupling which literally unlocks the full potential of the generally weak matter nonlinear response. By theoretically investigating the scattering of electromagnetic pulses by a thin nonlinear dispersive slab we show that a marked nonlinear pulse dynamics occurs only if the absolute value of the dielectric permittivity is very small over the pulse bandwidth. Such a nonlinear scenario stems from the fact that, in the ENZ regime, the field is spatially slowly varying and accordingly not characterized by a large number of "nodes" around which its amplitude is small, as the medium nonlinearity is allowed to uniformly affect the field over the entire bulk. In other words, as a viable strategy to attain a highly nonlinear response, here we suggest to enlarge the physical volume over which the nonlinearity is effective by means of the ENZ condition, as opposed to standard approaches that resort to field enhancement mechanisms or giant nonlinear parameters.

\section{Results}
\textbf{Pulse scattering by a slab supporting the ENZ regime}. The ENZ regime occurs when $|\epsilon(\omega)| \ll 1$ holds over the entire spectral bandwidth of the incident field. In conjunction with low absorption, this occurs if the field has a relatively narrow-band whose frequencies are very close to a zero of the real part of the medium permittivity. Permittivity zero-crossing-points are present both in metamaterials, where they can be tailored by choosing the relative metal-dielectric content of the unit cell, and in standard materials close to the resonant absorption frequencies. Here we analyze the second situation, and we model the dynamics of the matter polarization $\textbf{P}$ driven by the radiation electric field $\textbf{E}$ through the equation \cite{Kogaaa,Contii,Conti2,DiFal1,DiFal2}
\begin{equation} \label{polariz}
\frac{\partial^2 \textbf{P}}{\partial t^2}
        + \delta_e \omega_e \frac{\partial \textbf{P}}{\partial t}
        + \omega_e^2 \left( 1 + \frac{ \left| \textbf{P} \right|^2}{P_s^2} \right)^{-3/2} \textbf{P} = \epsilon_0 \left( \epsilon_s - 1 \right) \omega_e^2 \textbf{E}
\end{equation}
where $P_s$ is the saturation polarization that governs nonlinear oscillator behavior. For $|\textbf{P}|$ very much smaller than $P_s$, Eq.(\ref{polariz}) reduces to the single-pole Lorentz oscillator with resonant frequency $\omega_e$, loss coefficient $\delta_e \omega_e$ and static dielectric permittivity $\epsilon_s$. Therefore, in the linear regime, the dielectric permittivity experienced by monochromatic $e^{-i \omega t}$ fields is
\begin{equation} \label{eps}
\epsilon(\omega) = 1 + \frac{\epsilon_s -1}{1 - i \delta_e \left( \frac{\omega}{\omega_e} \right) - \left( \frac{\omega}{\omega_e} \right)^2},
\end{equation}
and it admits the zero-crossing-point
\begin{equation} \label{omega0}
\omega_0 =  \frac{\omega_e}{\sqrt{2}} \left\{  \left(\epsilon_s +1 - \delta_e^2 \right) +  \left[ \left( \epsilon_s +1 - \delta_e^2 \right)^2 - 4 \epsilon_s \right]^{1/2}  \right\}^{1/2}
\end{equation}
since $\textrm{Re} \left[ \epsilon \left( \omega_0 \right) \right] = 0$. For larger $|\textbf{P}|$ such that the condition $|\textbf{P}| \ll P_s$ still holds, Eq.(\ref{polariz}) describes a standard Kerr response \cite{Boyddd} whereas, at higher polarization strengths, it accounts for a realistic saturation of matter nonlinear response \cite{Janyan}. Therefore, Eq.(\ref{polariz}) contains all the ingredients necessary to describe a realistic and very general nonlinear wave-matter interaction in the ENZ regime.

\begin{figure*}
\center
\includegraphics[width=1\textwidth]{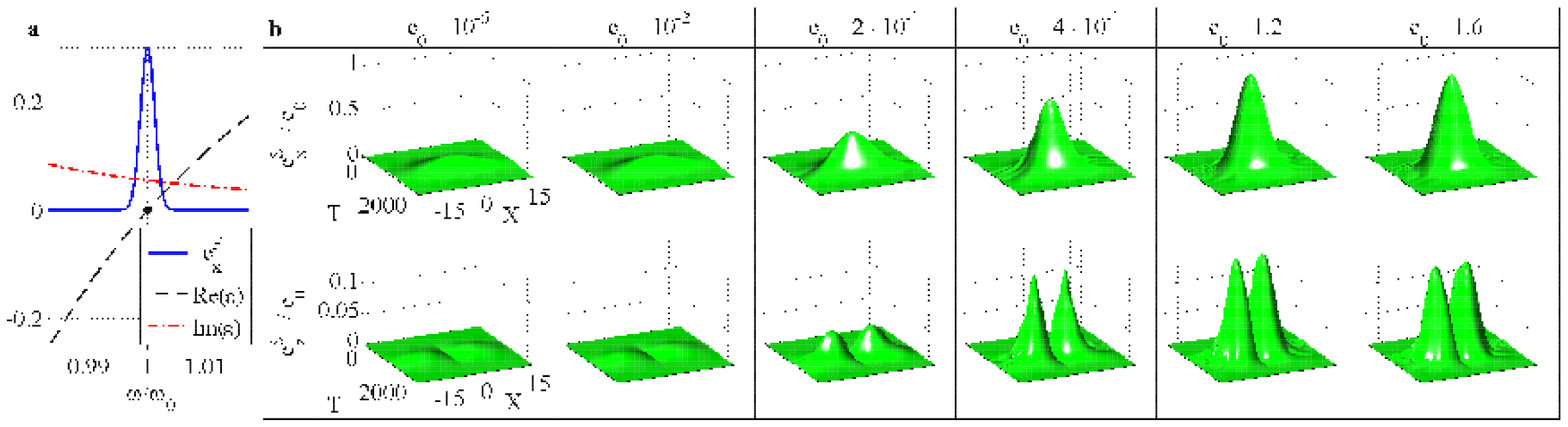}
\includegraphics[width=1\textwidth]{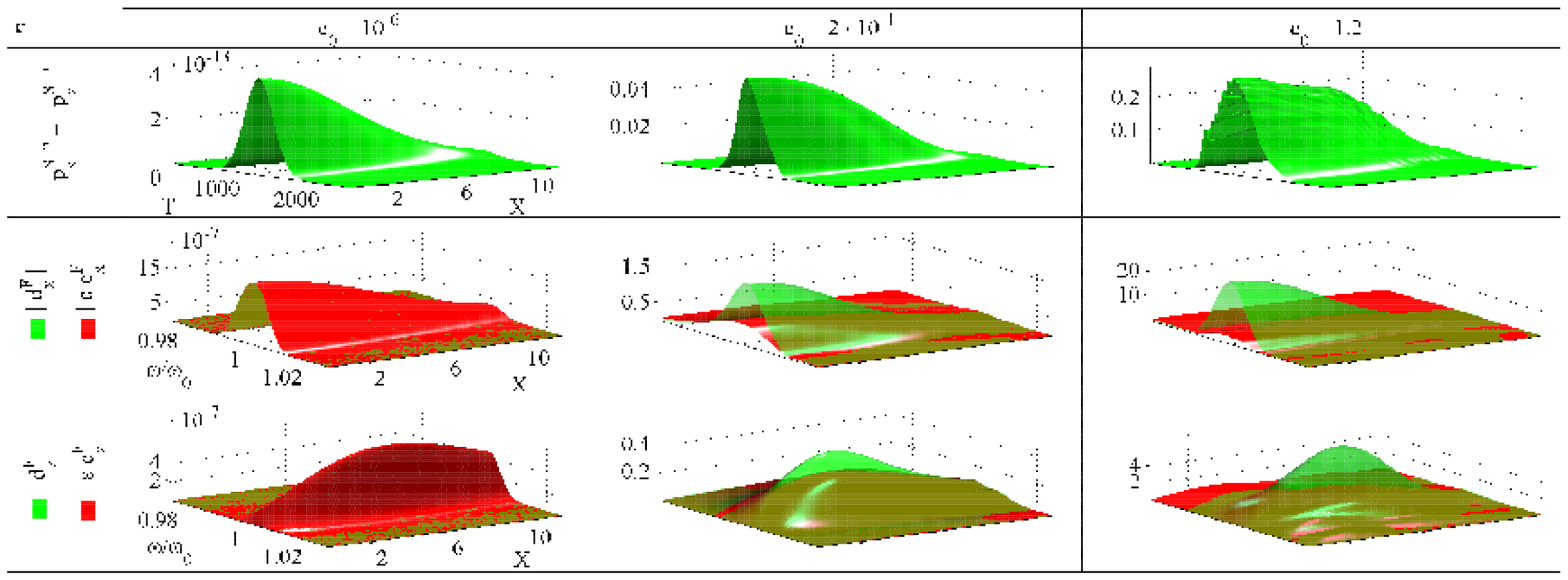}
\caption{\textbf{Nonlinear signatures of the scattering in the ENZ regime}, $\bar{\omega} = \omega_0$. (\textbf{a}) Spectrum $|e_x^F|$ of the incoming pulse (arbitrary units) and real and imaginary parts of the slab linear dielectric permittivity $\epsilon$. The interaction is in the ENZ regime since $|\epsilon| \ll 1$ holds over the pulse spectrum. (\textbf{b}) Absolute values of the dimensionless analytic signals, normalized to the amplitude of the incoming pulse $e_0$, $|e_x^S|/e_0$ and $|e_z^S|/e_0$ of the components of the outgoing pulse at the exit slab plane $Z_{out} = \omega_e L /(2c)$. The dependence of $|e_x^S|/e_0$ and $|e_z^S|/e_0$ on $e_0$ is the key signature of the nonlinear wave-matter interaction. (\textbf{c}) Square absolute values of the dimensionless analytic signal of the polarization $|p_x^S|^2 + |p_z^S|^2$, dimensionless displacement field components spectra $|d_x^F|$ and $|d_z^F|$ and dimensionless electric field components spectra multiplied by permittivity absolute value $|\epsilon e_x^F|$ and $|\epsilon e_z^F|$ at the slab middle plane $Z=0$. The occurrence of the nonlinear wave-matter interaction is further testified by two facts: 1) $|p_x^S|^2 + |p_z^S|^2$ is not very much smaller than one and 2) $|d_x^F| \neq |\epsilon e_x^F|$ and $|d_z^F| \neq |\epsilon e_z^F|$, i.e. the standard linear constitutive relation does not hold.}
\end{figure*}

In order to investigate such a regime we consider the scattering interaction reported in Fig.1a, where an electromagnetic pulse is launched from vacuum to orthogonally impinge on the surface of a dielectric slab whose polarization is described by Eq.(\ref{polariz}). The material dispersion parameters are $\delta_e = 0.01$ and $\epsilon_s = 1.2$ so that $\omega_0 = 1.095 \omega_e$ and the imaginary part of the permittivity around the zero-crossing-point is small, as reported in Fig.1b and in its inset, thus allowing the ENZ regime. We have set for the slab thickness $L = 1.25 \lambda_e$ where $\lambda_e = 2 \pi c / \omega_e$ is the resonant wavelength and therefore, since we will mainly consider pulses whose bandwidth is localized around the zero-crossing-point, the slab thickness is comparable with the main field wavelength. The pulse is a transverse magnetic (TM) excitation whose profile at the launching plane $
E_x (x,z_{in},t) = E_0 e^{-\frac{x^2}{\sigma^2}} e^{-\frac{(t-t_0)^2}{\tau^2}} \sin (\bar{\omega} t)$ has time shift $t_0 = 3.178 \cdot 10^3 \omega_e^{-1}$ and it is both spatially and temporally localized, $\sigma = 1.25 \lambda_e$ and $\tau = 1.059 \cdot 10^3 \omega_e^{-1}$ being its transverse and temporal widths, respectively. The amplitude $E_0$ and the carrier frequency $\bar{\omega}$ will be varied to investigate the ENZ regime. All the situations considered below are in the quasi-monochromatic regime since the carrier frequency $\bar{\omega}$ is always comparable with $\omega_e$ and the pulse spectral width $\delta \omega \simeq 1 / \tau = 9.442 \cdot 10^{-4} \omega_e$ is much smaller than $\omega_e$. Finite difference time domain simulations have been performed to solve Maxwell equations coupled with Eq.(\ref{polariz}) (see Methods). Results are reported using the dimensionless coordinates $(X,Z) = \omega_e (x,z) /c$, $T = \omega_e t$ and fields $(e_x,e_z) = \epsilon_0 (E_x,E_z)/P_s$, $h_y = H_y / (c P_s)$, $(p_x,p_z) = (P_x,P_z)/P_s$ which remove the resonant frequency $\omega_e$ and the saturation polarization $P_s$ from the model (see Methods).

\begin{figure*}
\center
\includegraphics[width=1\textwidth]{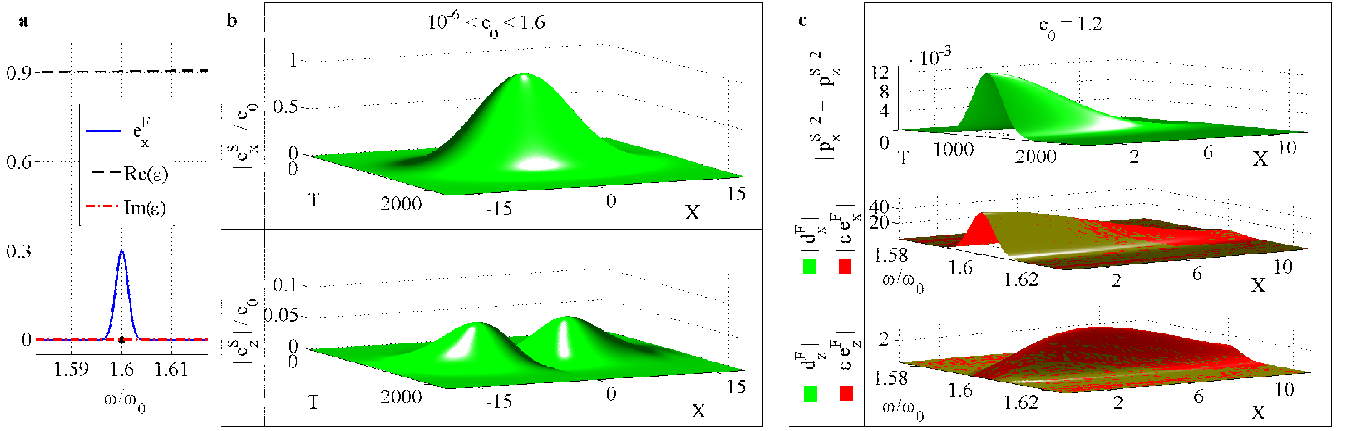}
\caption{\textbf{Linear character of the scattering outside of the ENZ regime}, $\bar{\omega} = 1.6 \omega_0$. (\textbf{a}) The incoming pulse spectrum $|e_x^F|$ is located at a spectral region where $|\epsilon| \simeq 0.9$ and therefore the interaction is outside of the ENZ regime. (\textbf{b}) For all the incoming pulses with amplitudes in the range $10^{-6} < e_0 < 1.6$ the absolute values of the normalized analytic signals $|e_x^S| / e_0$ and $|e_z^S| / e_0$ of the outgoing pulse are effectively identical. (\textbf{c}) For the pulse with amplitude $e_0 = 1.2$ the square absolute values $|p_x^S|^2 + |p_z^S|^2$ of the dimensionless analytic signal of the polarization within the slab are uniformly much smaller than $1$. For the same pulse amplitude the dimensionless displacement field components spectra, $|d_x^F|$ and $|d_z^F|$, coincide with the dimensionless electric field components spectra multiplied by permittivity absolute value, $|\epsilon e_x^F|$ and $|\epsilon e_z^F|$, within the slab. This results shows that pulse scattering outside of the ENZ regime is fully linear.}
\end{figure*}

\textbf{Highly nonlinear wave-matter interaction in the ENZ regime}. In the first set of simulations we have considered various pulses with carrier frequency located at the zero-crossing-point, i.e. $\bar{\omega} = \omega_0$, with different amplitudes $e_0 = \epsilon_0 E_0 / P_s$. In Fig.2a we have plotted the absolute value of the Fourier transform $e_x^F$ (arbitrary units) of the incoming pulse $e_x (0,Z_{in},T)$ superimposed to the real and imaginary parts of the permittivity. This figure clearly shows that $|\epsilon(\omega)|$ is much smaller than one all over the field spectrum, thus proving that the interaction occurs in the ENZ regime.

In Fig.2b, which contains the main result of the present paper, we report, for a number of different amplitudes $e_0$, the space-time profiles of the outgoing pulses. Specifically we plot the absolute values of the analytic signals $e_x^S(X,Z_{out},T)$ and $e_z^S(X,Z_{out},T)$ of the field components, normalized with respect to the amplitude $e_0$, at the exit slab surface $Z_{out} = \omega_e L /(2c)$. For $e_0 = 10^{-6}$ and $e_0 = 10^{-2}$ the normalized outgoing pulses are identical. This shows that the pulse amplitudes are small enough to prevent the matter nonlinearity from influencing the scattering. By increasing the pulse amplitude, for $e_0 = 2 \cdot 10^{-1}$ and $e_0 = 4 \cdot 10^{-1}$, the normalized outgoing pulses undergo dramatic transformation. The dependence on the input intensity is a key signature of the onset of the nonlinear regime. At higher pulse intensities, the normalized outgoing pulses having amplitudes $e_0 = 1.2$ and $e_0 = 1.6$ are almost identical, so that the nonlinear dependence of the scattering process on the input intensity saturates as expected from Eq.(\ref{polariz}).

In Fig.2c the transition from the linear regime to the nonlinear one is pictorially shown in two different and equivalent ways: (i) the analysis of the polarization of the slab, and (ii) the spectral analysis of the field dynamics within the slab. In the first row of Fig.2c we report the space-time profile of the polarization produced within the slab by some of the pulses considered in Fig.2b, by plotting the corresponding square absolute values $|p_x^S (X,0,T)|^2 + |p_z^S (X,0,T)|^2$ of the dimensionless analytic signal of the polarization at the slab middle plane $Z=0$. The relevance of such quantity stems from the fact that it is comparable with $\left| \textbf{P} \right|^2 / P_s^2$ which plays a key role in the linear-nonlinear transition of the wave-matter coupling (see Eq.(\ref{polariz})) through the nonlinear resonant frequency
\begin{equation} \label{nlome}
\bar{\omega}_{e}(\textbf{P}) = \omega_e \left( 1 + \frac{\left| \textbf{P} \right|^2}{P_s^2} \right)^{-3/4}.
\end{equation}
For $e_0 = 10^{-6}$ the normalized square polarization is so small that $\bar{\omega}_e^2 (\textbf{P}) \simeq \omega_e^2$. Accordingly, the linear regime (where the dielectric permittivity of Eq.(\ref{eps}) governs electromagnetic propagation) holds. For $e_0 = 2 \cdot 10^{-1}$ the normalized square polarization is small enough to allow a perturbative description of medium nonlinearity since $\bar{\omega}_e^2 (\textbf{P}) \simeq 1 - 3 \left| \textbf{P} \right|^2 / (2P_s^2)$. In this case Eq.(\ref{polariz}) is well-known to yield a nonlinear optical regime characterized by a Kerr nonlinearity \cite{Boyddd}. For $e_0 = 1.2$ the normalized square polarization is accordingly larger and does not generally allow a perturbative approximation of $\bar{\omega}_e^2 (\textbf{P})$, thus triggering the emergence of nonlinear saturation. In the second and third row of Fig.2c we plot the absolute values of the Fourier transforms $d_x^F(X,0,\omega)$ and $d_z^F(X,0,\omega)$ (green surfaces) of the components of the dimensionless displacement field $\textbf{d} = \textbf{e} + \textbf{p}$ at the slab middle plane $Z=0$. In the same rows we also plot the absolute values $|\epsilon(\omega) e_x^F(X,0,\omega)|$ and $|\epsilon(\omega) e_z^F(X,0,\omega)|$ (red surfaces) of the Fourier transforms of the electric fields components at the same plane $Z=0$ multiplied by the linear permittivity of Eq.(\ref{eps}). For $e_0 = 10^{-6}$ it is evident that $\textbf{d}^F = \epsilon \textbf{e}^F$ so that the standard linear relation between the displacement and the electric field holds. For $e_0 = 2 \cdot 10^{-1}$ and $e_0 = 1.2$ the discrepancy between the displacement field and its linear counterpart is dramatically evident thus restating that a marked nonlinear regime holds.

\begin{figure*}
\center
\includegraphics[width=1\textwidth]{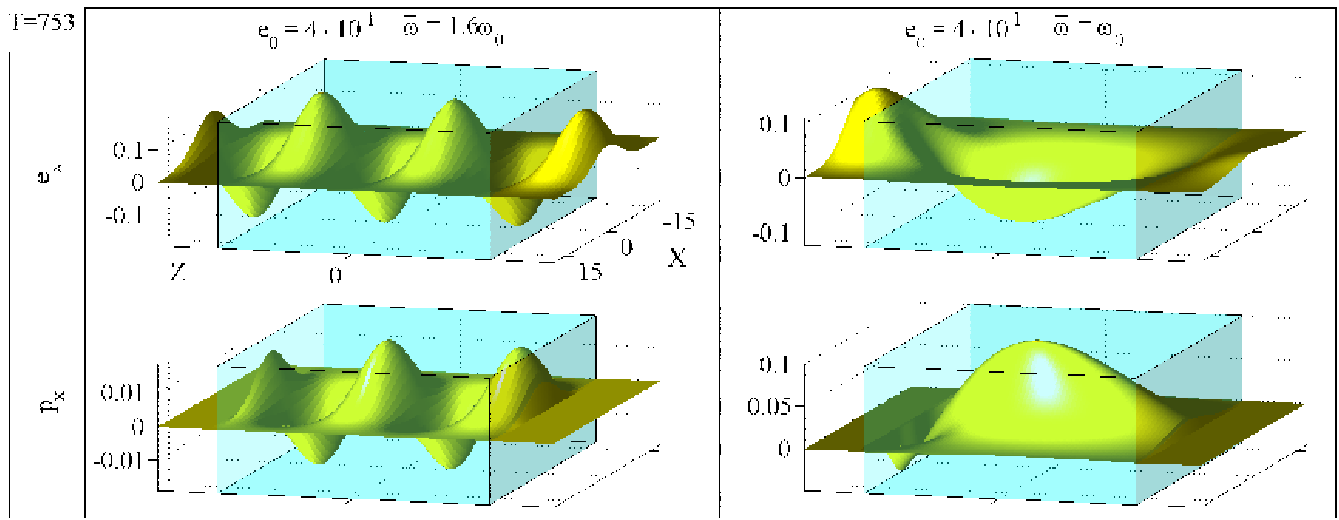}
\caption{\textbf{Mechanism supporting the full potential of the nonlinear wave-matter interaction in the ENZ regime}. Spatial profiles of the dimensionless electric $e_x$ and polarization $p_x$ fields at the time $T=753$ of the pulse with $e_0 = 0.4$ and $\bar{\omega} = 1.6 \omega_0$ (outside of the ENZ regime) and the pulse with $e_0 = 4 \cdot 10^{-1}$ and $\bar{\omega} = \omega_0$ (in the ENZ regime). In the first case the electric field $e_x$ is "large" only at the regions around the peaks of the wave and the time drift of such regions does not allow the polarization $p_x$ to increase and to trigger the nonlinear wave-matter coupling. In the second case, in a physically large volume, the spatially slowly varying character of $e_x$ yields the onset of the nonlinear regime. }
\end{figure*}

The most remarkable nonlinear trait of the scattering at hand in the ENZ regime is that the normalized amplitude of the outgoing pulse grows as the amplitude of the incident pulse is increased, as shown in Fig.2b. This phenomenology admits a simple explanation in terms of an effective nonlinear shift of the resonant frequency produced by the emergence of the nonlinear regime. From Fig.2c it is evident that the higher the amplitude $e_0$ of the incoming pulse the stronger the polarization induced within the slab. From Eq.(\ref{nlome}) this produces a local decrease of the nonlinear resonant frequency $\bar{\omega}_e$ which can be equivalently interpreted as an effective drift of the pulse spectrum toward the spectral region where the real part of the dielectric permittivity is positive (see Fig.1b). Therefore as $e_0$ increases the slab becomes more transparent to the pulse and allows larger energy transmission. Since the pulse is quasi-monochromatic and its spectrum is centered at the zero-crossing-point, even a slight spectral drift produces a relatively large change of the material response, which explains the dramatic dependence of the slab transmissivity on $e_0$ reported in Fig.2b.

It is worth noting that the interaction regime has a highly nonlinear character since a distinct pulse self-action occurs even if the slab is very thin, its thickness being comparable with the pulse carrier wavelength ($L = 1.368 \lambda_0$ where $\lambda_0 = 2 \pi c / \omega_0$ is the wavelength of the zero-crossing-point). In addition, we observe that the described nonlinear wave-matter interaction is not due to the enhancement of the longitudinal electric field component $e_z$ \cite{Campi2,Ciatt6,Ciatt7,Vince1,Ciatt8,Capret,Lukkkk}. Indeed, the transverse magnetic pulse incident from vacuum has a longitudinal component resulting from the finite size of the beam in the transverse direction (along the $x$-axis) and, due to the field matching at the slab edge, the longitudinal component within the slab is roughly $|\epsilon|^{-1} \simeq 20$ times larger than its vacuum counterpart. However, from the second and third row of Fig.2c, it evident that $|\epsilon e_x|$ and $|\epsilon e_z|$ are comparable within the slab so that $|e_x|$ and $|e_z|$ are comparable as well. This demonstrates that in this case the longitudinal field component $e_z$ does not play the usual leading role we have grown accustomed to,
notwithstanding the fact that the longitudinal component is still enhanced. As a consequence, the same nonlinear mechanism is triggered with incident transverse electric pulses. Therefore, a distinct advantage one gains is that, by focusing more tightly an incident transverse magnetic pulse, both enhancement of the longitudinal field and the kind of nonlinear behavior we have just discussed can in principle combine for optimal results.

\textbf{Absence of nonlinear effects outside of the ENZ regime}. The remarkable impact of matter nonlinearity on electromagnetic propagation in the ENZ regime is rendered even more peculiar and fascinating by the fact that no deviation from linear behavior occurs in the pulse scattering outside of the ENZ regime. This important fact is proved by a second set of simulations where we have considered various pulses identical to those discussed in Fig.2, except for their carrier frequencies, which we have set to $\bar{\omega} = 1.6  \omega_0$. The results are reported in Fig.3, where we have plotted the same quantities considered in Fig.2. Figure 3a clearly shows that the incoming pulse spectrum $|e_x^F|$ is located in a spectral region where $|\epsilon| \simeq 0.9$. Therefore, interaction does not occur in the ENZ regime, and the slab shows purely linear dielectric behavior. Figure 3b shows that the pulse amplitude $e_0$ has no impact on the scattering since for all the considered pulses with $10^{-6} < e_0 < 1.6$ the profiles of the absolute values of the normalized analytic signals $|e_x^S| / e_0$ and $|e_z^S| / e_0$ are all the same. Figure 3c reveals that the square absolute values $|p_x^S|^2 + |p_z^S|^2$ of the dimensionless analytic signal of the polarization within the slab, even for the relatively large pulse amplitude $e_0 = 1.2$, is uniformly much smaller than $1$. In addition, Fig. 3c shows that, for the same large pulse amplitude $e_0 = 1.2$, the standard linear relation $\textbf{d}^F = \epsilon \textbf{e}^F$ holds between the displacement and the electric field within the slab. As a result, pulse scattering does not reveal any nonlinear traits if the pulse carrier frequency is not close to the zero-crossing-point, no matter how large the pulse amplitude $e_0$ is made. We conclude that the slab governed by Eq.(\ref{polariz}) does not play host to a nonlinear wave-matter interaction, unless the ENZ regime is exploited.

\textbf{Mechanism supporting the highly nonlinear wave-matter coupling in the ENZ regime}. The discussed nonlinear wave-matter interaction that occurs only in the ENZ regime admits an explanation which is closely related to the slowly varying character of the electromagnetic field imposed by the very small permittivity. In Fig.4 we plot the spatial profiles of the dimensionless electric $e_x$ and polarization $p_x$ fields, at a specific time $T = 753$, for two of the pulses considered above, and having the same amplitude $e_0 = 4 \cdot 10^{-1}$ and different carrier frequencies $\bar{\omega} = 1.6 \omega_0$ and $\bar{\omega} = \omega_0$. The first pulse (left column of Fig.4) is tuned outside of the ENZ regime. Its scattering is purely linear, whereas the second pulse (right column of Fig.4) with $\bar{\omega} = \omega_0$ is in the ENZ regime and it has been shown to display a marked nonlinear dynamics. Consider the first pulse with $\bar{\omega} = 1.6 \omega_0$ and note that, at the beginning of the scattering interaction, pulse propagation is evidently linear since the initial electric field is small. Therefore the profile of the pulse electric field has a considerable number of nodes, the regions of low electric field around such nodes drift in time due to propagation, the amplitude $| {\bf A} ({\bf r},t)|$ of the electric field ${\bf E}({\bf r},t) = Re \left[  {\bf A}({\bf r},t) e^{-i 1.6 \omega_0 t} \right]$ driving the polarization at each point rapidly varies between zero and it maximum value. Since in the linear regime Eq.(\ref{polariz}) yields
\begin{equation}
\textbf{P}(t) = \epsilon_0 \int_{-\infty}^{t} dt' \chi(t-t') \textbf{E}(t'),
\end{equation}
where $\chi(t)$ is the Lorentz susceptibility, we conclude that the rapid variation of the electric field amplitude forbids the polarization from locally increasing and from driving the scattering process out of the linear regime. In the case of the second pulse with $\bar{\omega} =  \omega_0$ the initial linear dynamics forces the electric field to be spatially slowly varying within the bulk since it is in the ENZ regime. As a consequence, very few nodes appear in the electric field profile and correspondingly the amplitude $| {\bf A} ({\bf r},t)|$ of the electric field ${\bf E}({\bf r},t) = Re \left[ {\bf A} ({\bf r},t) e^{-i \omega_0 t} \right]$ has a time variation scale much slower than the first pulse, occuring over a physically large portion of the bulk. Hence the polarization is efficiently pumped by the electric field (much more than the pulse tuned outside of the ENZ regime) so that it correspondingly increases to the point of triggering the onset of the nonlinear wave-matter coupling we have discussed above.

\section{Discussion}
In conclusion we have shown that a novel and highly nonlinear wave-matter coupling occurs if the medium's linear permittivity is very small over the entire electromagnetic field bandwidth. The realization of the full potential offered by this nonlinear matter response is due to the spatially slowly varying character of the electromagnetic field in the ENZ regime. In fact, the ENZ regime produces more efficient coupling between the electric field and the medium polarization field by enlarging the effective portion of the bulk hosting the nonlinear interaction. Unlike most strategies proposed to date in literature in order to achieve a highly nonlinear response, the nonlinear regime we have discussed here is not triggered by either cavity effects or by large nonlinear coefficients. The strategy we have outlined is simpler and more suitable to be exploited in a number a different configurations since it requires only that the field's main frequency coincides with a zero-crossing-point of the real part of the dielectric permittivity, preferably at the crossing point where absorption may be neglected. In addition, in view of such flexibility, the nonlinear coupling we have discussed may be triggered even in the presence of other mechanisms that are known to enhance matter-wave coupling, thus likely to produce additional hitherto unknown, highly nonlinear effects. For example, the enhancement of the field component normal to the surface between vacuum and the ENZ material may be triggered in the scattering experiment considered in this paper simply by confining more tightly the incoming pulse along the transverse direction. The simplicity, generality and flexibility properties of the discussed highly nonlinear regime make it an ideal platform for conceiving a number of applications and a novel generation of ultra-compact and fast devices for manipulating light.

\section{Methods}
\textbf{Full-wave simulations}. Electromagnetic propagation within the slab is described by Maxwell equations coupled to Eq.(\ref{polariz}) for the polarization so that, by using the dimensionless coordinates ${\bf R} = \omega_e {\bf r}/c$, $T = \omega_e t$ and dimensionless fields ${\bf e} = \epsilon_0 {\bf E}/P_s$, ${\bf h} = {\bf H}/ (c P_s)$, ${\bf p} = {\bf P}/P_s$, the pulse scattering is described by the set of equations
\begin{eqnarray}
\nabla_{\bf R} \times {\bf e} &=& -\frac{\partial {\bf h}}{\partial T}, \nonumber \\
\nabla_{\bf R} \times {\bf h} &=& \frac{\partial {\bf e}}{\partial T} + \frac{\partial {\bf p}}{\partial T}, \nonumber \\
\frac{\partial^2 \textbf{p}}{\partial T^2}
        + \delta_e  \frac{\partial \textbf{p}}{\partial T}
        + \left( 1 + \left| \textbf{p} \right|^2 \right)^{-3/2} \textbf{p} &=& \left( \epsilon_s - 1 \right) \textbf{e}
\end{eqnarray}
within the slab, by the first two equations with $\textbf{p} = \textbf{0}$ outside of the slab and matching conditions of the tangential electric and magnetic field components at the slab surfaces.

The pulse scattering by the slab both in and outside of the ENZ regime was simulated by solving the above equations through a home-made finite-difference time-domain code suitable for dealing with transverse magnetic pulses ${\bf e}(X,Z,T) = e_x (X,Z,T) {\bf i} + e_z (X,Z,T) {\bf k}$, ${\bf h}(X,Z,T) = h_y (X,Z,T) {\bf j}$. At the edges of the computational domain perpendicular to the $Z$ axis scattering boundary condition where adopted whereas at those perpendicular to the $X$ axis the vanishing of all the field components was imposed.

\section{Acknowledgements}

A.C. and C.R. acknowledge support from U.S. Army International Technology Center Atlantic for financial support (Grant No. W911NF-14-1-0315). A.D.F. acknowledges support from EPSRC (EP/I004602/1). D.F. acknowledges support from the European Research Council under the European Union’s Seventh Framework Programme (FP/2007-2013)/ERC GA 306559.

\section{Author contributions}
A.C. and C.R. developed the concept and performed the numerical simulations. A.M., A.D.F., D.F. and M.S. contributed to the interpretation of the results of the numerical simulations. All authors contributed to discussions and to writing of the paper.


\end{document}